\documentclass[a4paper]{article}

\usepackage{graphicx,cite}

\newcommand{\RR}{\mathbf{R}}
\newcommand{\ZZ}{\mathbf{Z}}

\newtheorem{theorem}{Theorem}
\newtheorem{prop}[theorem]{Proposition}

\begin{document}

\title{On the spectrum of a waveguide with periodic cracks}

\author{Konstantin Pankrashkin\\[\bigskipamount]
Laboratoire de math\'ematiques d'Orsay (CNRS UMR 8628)\\
Universit\'e Paris-Sud, B\^atiment 425\\
91400 Orsay, France}

\date{\it Dedicated to the memory of Pierre Duclos}

\maketitle

\begin{abstract}
\noindent The spectral problem on a periodic domain with cracks is studied.
An asymptotic form of dispersion relations is calculated under assumption that
the opening of the cracks is small.

\end{abstract}

Let $\Omega\subset \RR^2$ be a connected smooth domain satisfying the following
conditions:
\begin{itemize}
\item $\Omega$ is invariant under the shift $(x,y)\mapsto (x+1,y)$,
\item the domains $\Omega_n:=\Omega\cap \Big( [n,n+1]\times\RR\Big)$, $n\in\ZZ$, are bounded,
\item the points $A_n:=(n,0)$  are interior points of $\Omega$.
\end{itemize}
Assume that the vertical lines $L_n=(n,\RR)$, $n\in\ZZ$, are non-tangent to the boundary of $\Omega$
and denote $\Gamma_n^\varepsilon:=L_n\cap \Omega \setminus B_\varepsilon(A_n)$, where $B_\varepsilon(A_n)$
is the ball of radius $\varepsilon$ centered at $A_n$.

\begin{figure}

\begin{center}

\includegraphics[height=50mm]{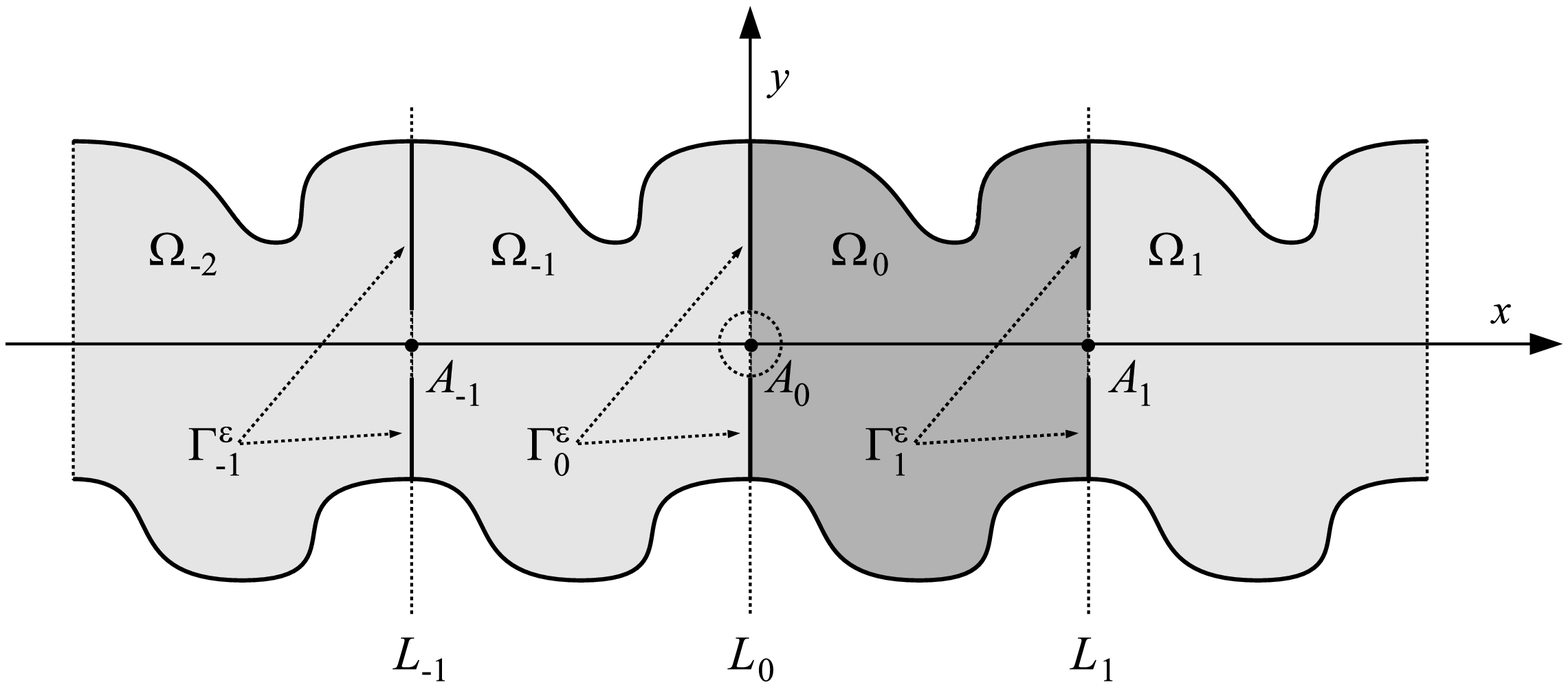}

\title{Figure 1. Domain $\Omega$}

\end{center}

\end{figure}

Consider in $L^2(\Omega)$ the operator $H_\varepsilon$ which is the Laplacian with the Neumann boundary conditions
on $\partial\Omega$ and on all the curves $\Gamma_n^\varepsilon$.
We are studying some spectral properties of $H_\varepsilon$ as $\varepsilon$ tends to $0$.

The operator $H_\varepsilon$ can be viewed as the Hamiltonian of elementary cells $\Omega_n$
coupled by small windows. For positive $\varepsilon$ one has a periodic system with a band spectrum, while
at $\varepsilon=0$ the system is decoupled. There is a number of papers concerning the spectral asymptotics
for waveguides with windows or cracks in various settings, see e.g. \cite{DH,EV,gdl,popov,Y1,Yper}.
In particular, the paper \cite{Yper} studied a situation which is close to ours, more precisely,
the case of the Dirichlet boundary conditions on a similar periodic structure with particular symmetries (straight strip with symmetric cuts),
and the asymptotics of the lowest dispersion relation was calculated.
All the papers cited use the method of matching of asymptotic expansions \cite{il}, which is quite sensible
to the geometric properties. We give here another proof based on the analysis of operator pencils
involving boundary integrals and inspired by the recent monograph \cite{Amm}. This allows one to obtain the asymptotics of the spectrum in
rather general situations (the assumption made above can be relaxed is various directions, we have just chosen
a basic situation in order to keep simple notations). Integral operator pencils of a similar type were used by Pierre Duclos with
co-authors as a part of the skeleton method for the study of multi-particle problems\cite{CHR}.

By periodicity, the spectrum of $H_\varepsilon$ can be studied with the help of  the Bloch theory.
For $\theta\in[0,2\pi]$ consider the values $E_n(\theta;\varepsilon)$ for which there exists
a non-zero function $\Psi_n(x,y;\theta,\varepsilon)$ satisfying the equation $-\Delta \Psi_n(x,y;\theta,\varepsilon)=E_n(\theta,\varepsilon)\Psi_n(x,y;\theta,\varepsilon)$, the above
boundary conditions, and the Bloch quasiperiodicity condition $\Psi_n(x+1,y;\theta,\varepsilon)=
e^{i\theta}\Psi_n(x,y;\theta,\varepsilon)$; here $\theta\in[0,2\pi]$ is a real number called quasimomentum.
By a suitable reordering 
one can assume that the functions $\theta\to E_n(\theta,\varepsilon)$ are continuous
and $2\pi$-periodic; these function are called dispersion relations of $H_\varepsilon$.
The image of a dispersion relation is called a spectral band of $H_\varepsilon$, and the spectrum
of $H_\varepsilon$ is the union of the spectral bands.

We are interested in the asymptotic form of the dispersion relations for small $\varepsilon$.
By the standard variational arguments, see e.g. Proposition 1.3 in \cite{Yper}
or proposition 9.8 in \cite{DH}, each function $E_{n}(\cdot,\varepsilon)$ converges to a constant function equal to an eigenvalue
of the decoupled system, i.e. to an eigenvalue of the Neumann Laplacian on $\Omega_0$.
Our main result is the following theorem. 

\begin{theorem}\label{thm1}
Let $E$ be a simple eigenvalue of the Neumann Laplacian in $\Omega_0$
and $u$ be the corresponding normalized eigenfunction.
As $\varepsilon\to 0$, the dispersion relation of $H_\varepsilon$
near $E$ is of the form
\[
E(\theta,\varepsilon)=E+\frac{2\pi}{|\log \varepsilon|} \big|\,u(A_0)-e^{i\theta} u(A_1)\big|^2
+O\Big(
\frac{1}{\log^2 \varepsilon}
\Big)
\]
uniformly in $\theta\in[0,2\pi]$. 
\end{theorem}
We note that $E$ can be \emph{any} simple eigenvalue, and not just the bottom one.

The rest of the paper is devoted to the proof of theorem \ref{thm1}.

Let $G_n(r,r';z)$ be the Green function of the Neumann Laplacian $N_n$ in $\Omega_n$, i.e. the integral kernel of the resolvent
$(N_n-z)^{-1}$. Recall that, if $z$ is not an eigenvalue of $N_n$, then the boundary value problem
\[
(-\Delta-z) f =0 \mbox{ in } \Omega_n,
\quad
\frac{\partial f}{\partial \nu}\Big|_{\partial\Omega_n} =g,
\]
where $\nu$ is the exterior normal vector, has the unique solution
\[
f(r)=\int_{\partial \Omega_n} G_n(r, r';z) g(r')\, dl(r').
\]

Consider a Bloch solution $\Psi$ of $H_\varepsilon$ corresponding a quasimomentum $\theta$
and to an eigenvalue $z=E(\theta)$. Introduce the functions
\[
f_n(y):= \frac{\partial \Psi (x,y)}{\partial x}\Big|_{x=n}, \quad y\in(-\varepsilon,\varepsilon).
\]
Denote by $\Psi_n$ the restriction of $\Psi$ to $\Omega_n$. One has, obviously,
\[
\Psi_n(x,y)=-\int_{-\varepsilon}^\varepsilon
G_n(x,y;n,y';z) f_n(y') dy'
+\int_{-\varepsilon}^\varepsilon
G_n(x,y;n+1,y';z) f_{n+1}(y') dy'.
\]
The Bloch condition for $\Psi$ implies $f_{n+1}(y')=e^{i\theta} f_n(y')$.
The continuity of $\Psi$ at $x=n$ takes the form $\Psi_n(n,y)=\Psi_{n-1}(n,y)$,
or, using the above integral representation,
\begin{eqnarray*}
e^{i\theta} \int_{-\varepsilon}^{\varepsilon} G_n(n,y;n+1,y';z)f_n(y')dy'-\int_{-\varepsilon}^{\varepsilon} G_n(n,y;n,y';z)f_n(y')dy'\\
{}=\int_{-\varepsilon}^{\varepsilon} G_{n-1}(n,y;n,y';z)f_n(y')dy' -e^{-i\theta}\int_{-\varepsilon}^{\varepsilon} G_{n-1}(n,y;n-1,y';z)f_n(y')dy'.
\end{eqnarray*}
Using the obvious identity $G_n(x,y;x',y';z)=G_0(x-n,y;x'-n,y';z)$ one arrives at a single integral equation
\begin{equation}
 \label{eq-kte}
 K_{\theta,\varepsilon}(z) f_{\theta,\varepsilon}:= \int_{-\varepsilon}^\varepsilon
 K_{\theta,\varepsilon} (y,y';z) f_{\theta,\varepsilon}(y')\,dy'=0, \quad f_{\theta,\varepsilon}:=f_0,
\end{equation}
with the integral kernel
\begin{eqnarray*}
 K_{\theta,\varepsilon} (y,y';z) =G_0(0,y;0,y';z)+G_0(1,y;1,y';z)\\
 \qquad {}- e^{i\theta} G_0(0,y;1,y';z) -  e^{-i\theta} G_0(1,y;0,y';z).
\end{eqnarray*}
In order to determine the dispersion relations of $H_\varepsilon$
one needs to find the values of $z$ for which the equation (\ref{eq-kte})
has non-zero solutions, i.e. the nonlinear eigenvalues of $K_{\theta,\varepsilon}$.

As noted above, integral equations of this type were studied in details in the recent monograph \cite{Amm},
and here we add some details needed for the treatment of the periodic problem.
First of all, by Lemma 5.6 in \cite{Amm}, for each fixed $\theta$, the equation (\ref{eq-kte})
has a unique solution in a neighborhood of $E$ if $\varepsilon$ is small enough.

Like in \cite{Amm} introduce the Hilbert space
\[
X_\varepsilon:=\big\{
\varphi:\,\|\varphi\|^2_{X_\varepsilon}:=\int_{-\varepsilon}^{\varepsilon} \sqrt{\varepsilon^2-y^2} |\varphi(y)|^2\,dy<+\infty
\big\},
\] 
and the linear space
\[
Y_\varepsilon:=\{
\psi\in C[-\varepsilon,\varepsilon]: \psi'\in X_\varepsilon\}.
\]
By the trace theorems, the functions $f_{\theta,\varepsilon}$ belong to both $X_\varepsilon$ and $Y_\varepsilon$.

We emphasize that the symbol $\langle \cdot, \cdot\rangle$ will \emph{always} denote the usual scalar product
in $L^2[-\varepsilon,\varepsilon]$, i.e.
\[
\langle f, g\rangle:=
\int_{-\varepsilon}^\varepsilon \overline{f(y)} \,g(y)\, dy.
\]

Using Lemma 5.1 in \cite{Amm} and the spectral  representation of the Green function $G_0$
one can decompose $K_{\theta,\varepsilon}$ as follows:
\begin{equation}
         \label{eq-decomp}
K_{\theta,\varepsilon}(z):=-\frac{1}{2\pi} L_\varepsilon
+ \frac{M_{\theta,\varepsilon}}{E-z} +R_{\theta,\varepsilon}(z),
\end{equation}
where $L_\varepsilon$ is the integral operator
\[
L_\varepsilon f(x)=\int_{-\varepsilon}^{\varepsilon}
\log |x-y| \,f(y)\, dy,
\]
$M_{\theta,\varepsilon}$ is the rank one operator
\[
M_{\theta,\varepsilon} f (x) =u_{\theta,\varepsilon}(x) \int_{-\varepsilon} \overline{u_{\theta,\varepsilon}(y)}\, f(y)\,dy,
\quad
u_{\theta,\varepsilon}(y):=u(0,y)-e^{i\theta} u(1,y),
\]
and $R_{\theta,\varepsilon}(z)$ is an integral operator
\[
R_{\theta,\varepsilon}(z) f(x)=\int_{-\varepsilon}^{\varepsilon}
R_{\theta,\varepsilon}(x,y;z)\,f(y)\, dy,
\]
where the kernel $R_{\theta,\varepsilon}(x,y;z)$ is holomorph in $(z,\theta)$
and of H\"older class $C^{1+\alpha}$ with respect to $(x,y)$; here $\alpha\in(0,1)$
is arbitrary. 
Note that the functions $u_{\theta,\varepsilon}$ are in both $X_\varepsilon$ and $Y_\varepsilon$ as well.

By Lemma 5.2 in \cite{Amm}, the operator $L_\varepsilon:X_\varepsilon\to Y_\varepsilon$
is a bijection for $\varepsilon$ small enough, and, by Lemma 5.4 in \cite{Amm}, one has,
uniformly in $\theta\in[0,2\pi]$ and $z$ in a neighborhhod of $E$, the norm estimate
\begin{equation}
  \label{eq-oeps}
\|L^{-1}_\varepsilon R_{\theta,\varepsilon}(z)\|_{L(X_\varepsilon,X_\varepsilon)}= O \Big(
\frac{1}{\log \varepsilon}
\Big);
\end{equation}
here and below $L(X_\varepsilon,X_\varepsilon)$ is the Banach space of bounded linear operators on $X_\varepsilon$. 
We will need the following estimates:

\begin{prop}\label{prop2} As $\varepsilon\to 0$ one has, uniformly in $\theta$,
\begin{equation} \label{eq-est1}
\langle L^{-1}_\varepsilon
u_{\theta,\varepsilon},u_{\theta,\varepsilon}\rangle= \frac{\big|u_{\theta,\varepsilon}(0)\big|^2}{\log\varepsilon}
+ O\Big(\frac{1}{|\log\varepsilon|^2}\Big).
\end{equation}
Futhermore, there exists $A>0$ with the following property: If $B$ is a bounded operator in 
$X_\varepsilon$ then, as $\varepsilon\to 0$, one has
\begin{equation} \label{eq-est2}
\big|\langle B L^{-1}_\varepsilon
u_{\theta,\varepsilon},u_{\theta,\varepsilon}\rangle\big| \le \frac{A\|B\|_{L(X_\varepsilon,X_\varepsilon)}}{|\log \varepsilon|}.
\end{equation}

\end{prop}

\noindent{\bf Proof}
As in Lemma 5.4 of \cite{Amm} we use an explicit form for the inverse of $L_\varepsilon$
and some properties of the finite Hilbert transform. We have
\begin{eqnarray*}
L_\varepsilon^{-1}u_{\theta,\varepsilon}(x)=
-\frac{1}{\pi^2\sqrt{\varepsilon^2-x^2}} \int_{-\varepsilon}^\varepsilon \frac{\sqrt{\varepsilon^2-y^2}\, u'_{\theta,\varepsilon}(y)}{x-y}dy\\
{}
+\frac{a(u_{\theta,\varepsilon})}{\pi \log \frac{\varepsilon}{2}} \cdot \frac{1}{\sqrt{\varepsilon^2-x^2}}=: I_1+I_2,
\end{eqnarray*}
where
\[
a(u_{\theta,\varepsilon})=u_{\theta,\varepsilon}(x)-L_\varepsilon v_{\theta,\varepsilon}(x)
\]
(the choice of $x$ is arbitrary) 
with
\[
v_{\theta,\varepsilon}(x)=-\frac{1}{\pi^2\sqrt{\varepsilon^2-x^2}}
\int_{-\varepsilon}^\varepsilon \frac{\sqrt{\varepsilon^2-y^2}\, u'_{\theta,\varepsilon}(y)}{x-y} dy;
\]
here and below all the integrals are understood in the sence of the Cauchy principal value.

We will use the following well known estimate: for any $\alpha\in(0,1)$ there exists $C>0$ such that
for any $\varphi\in C^\alpha[-1,1]$  and $x\in[-1,1]$ there holds
\[
\Big| \int_{-1}^1\frac{\varphi(y)}{x-y}dy
\Big|\le C \|\varphi\|_{C^{\alpha}[-1,1]}.
\]

One has then
\begin{eqnarray}
\|
I_1
\|_{X_\varepsilon}\nonumber
=\frac{1}{\pi^2}\,\sqrt{
\int_{-\varepsilon}^\varepsilon \frac{1}{\sqrt{\varepsilon^2-x^2}}
\Big(
\int_{-\varepsilon}^\varepsilon
\frac{\sqrt{\varepsilon^2-y^2} u'_{\theta,\varepsilon}(y)}{x-y}dy
\Big)^2
dx\qquad
}\nonumber\\
\le
\frac{1}{\pi^2}\, 
\sqrt{\int_{-\varepsilon}^\varepsilon \frac{1}{\sqrt{\varepsilon^2-x^2}} dx}\,\,\,
\Big\|
\int_{-\varepsilon}^\varepsilon
\frac{\sqrt{\varepsilon^2-y^2} u'_{\theta,\varepsilon}(y)}{x-y}\,dy\Big\|_{L^\infty[-\varepsilon,\varepsilon]}\qquad\nonumber\\
=
\frac{\varepsilon}{\pi^{3/2}} \Big\|
\int_{-1}^1
\frac{\sqrt{1-y^2}\, u'_{\theta,\varepsilon}(\varepsilon y)}{x-y}\,dy\Big\|_{L^\infty[-1,1]}\nonumber\\
\le 
\frac{C \varepsilon}{\pi^{3/2}}\, \Big\| \sqrt{1-y^2}\, u'_{\theta,\varepsilon}(\varepsilon y) \Big\|_{C^\alpha[-1,1]}, 
          \label{holder}
\end{eqnarray}
and one obtains $\|I_1\|_{X_\varepsilon}=O(\varepsilon)$ uniformly in $\theta$.
This implies $\langle I_1,u_{\theta,\varepsilon}\rangle=O(\varepsilon)$ as well.
Hence it is sufficient to study the asymptotics of $\langle I_2,u_{\theta,\varepsilon}\rangle$.
We have
\begin{eqnarray*}
\langle I_2,u_{\theta,\varepsilon}\rangle=
\frac{1}{\pi \log |\varepsilon/2|}\bigg(
\int_{-\varepsilon}^\varepsilon \frac{\big|u_{\theta,\varepsilon}(x)\big|^2}{\sqrt{\varepsilon^2-x^2}}dx-
\int_{-\varepsilon}^\varepsilon \overline{L_\varepsilon v_{\theta,\varepsilon}(y)}\,\displaystyle\frac{u_{\theta,\varepsilon}(y)}{\sqrt{\varepsilon^2-y^2}}\,
dy
\bigg)\\
=:\frac{1}{\pi \log |\varepsilon/2|}\,(K_1-K_2).
\end{eqnarray*}
There holds
\begin{eqnarray*}
K_1=\int_{-\varepsilon}^\varepsilon \frac{\big|u_{\theta,\varepsilon}(x)\big|^2}{\sqrt{\varepsilon^2-x^2}}dx=
\int_{-1}^1 \frac{\big|u_{\theta,\varepsilon}(\varepsilon x)\big|^2}{\sqrt{1-x^2}}dx \qquad\qquad\qquad\qquad\\
=\big|u_{\theta,\varepsilon}(0)\big|^2\,\int_{-1}^1 \frac{dx}{\sqrt{1-x^2}}
+\int_{-1}^1 \frac{\big|u_{\theta,\varepsilon}(\varepsilon x)\big|^2-\big|u_{\theta,\varepsilon}(0)\big|^2}{\sqrt{1-x^2}}dx\\
=\pi \big|u_{\theta,\varepsilon}(0)\big|^2 + O(\varepsilon),
\end{eqnarray*}
where the remainder term
is uniform in $\theta$.

To estimate $K_2$ let us note first that, by the same arguments as in (\ref{holder}),
\[
\int_{-\varepsilon}^\varepsilon
\frac{\sqrt{\varepsilon^2-y^2}}{x-y} u_{\theta,\varepsilon}'(y)\,dy=:K_3(x)=O(\varepsilon)
\]
uniformly in $x$ and $\theta$.
Therefore,
\begin{eqnarray*}
\sup_{x\in[-\varepsilon,\varepsilon]}\big|L_\varepsilon v_{\theta,\varepsilon}(x)\big|
=\frac{1}{\pi^2}\,\sup_{x\in[-\varepsilon,\varepsilon]}\Big|\int_{-\varepsilon}^\varepsilon \log|x-y|\cdot K_3(y)\cdot
\frac{1}{\sqrt{\varepsilon^2-y^2}}\, dy\Big| \\
\le \frac{1}{\pi^2}\,\sup_{y\in[-\varepsilon,\varepsilon]}
\big|K_3(y)\big|
\sup_{x\in[-\varepsilon,\varepsilon]}\int_{-\varepsilon}^\varepsilon \frac{\Big|\log|x-y|\Big|}{\sqrt{\varepsilon^2-y^2}} dy\\
=\frac{1}{\pi}\,\sup_{y\in[-\varepsilon,\varepsilon]}
|K_3(y)|\, \log \frac{\varepsilon}{2} 
=O(\varepsilon \log\varepsilon)
\end{eqnarray*}
uniformly in $\theta$. Finally,
\[
|K_2|\le
\sup_{x\in[-\varepsilon,\varepsilon]}\big|L_\varepsilon v_{\theta,\varepsilon}(x)\big|
\int_{-\varepsilon}^\varepsilon
\displaystyle\frac{\big|u_{\theta,\varepsilon}(y)\big|}{\sqrt{\varepsilon^2-y^2}}\,
dy=O(\varepsilon\log \varepsilon).
\]
We obtain
\begin{eqnarray*}
\langle L^{-1}_\varepsilon
u_{\theta,\varepsilon},u_{\theta,\varepsilon}\rangle=
\langle I_1,u_{\theta,\varepsilon}\rangle +\langle I_2,u_{\theta,\varepsilon}\rangle
=\langle I_1,u_{\theta,\varepsilon}\rangle +\frac{K_1}{\pi\log \frac{\varepsilon}{2}}-
\frac{K_2}{\pi\log \frac{\varepsilon}{2}}\\
=O(\varepsilon) + 
\frac{\pi \big|u_{\theta,\varepsilon}(0)\big|^2 + O(\varepsilon)}{\pi\log \frac{\varepsilon}{2}}
+\frac{O(\varepsilon\log \varepsilon)}{\pi\log \frac{\varepsilon}{2}}
=\frac{\big|u_{\theta,\varepsilon}(0)\big|^2}{\log \varepsilon}
+O\Big(\frac{1}{|\log\varepsilon|^2}\Big),
\end{eqnarray*}
which proves (\ref{eq-est1}).

To show (\ref{eq-est2}) we use first the Cauchy-Schwartz inequality for $X_\varepsilon$,
\begin{eqnarray*}
\big|\langle  B L^{-1}_\varepsilon u_{\theta,\varepsilon},u_{\theta,\varepsilon}\rangle\big|
=\Big|\int_{-\varepsilon}^\varepsilon \sqrt{\varepsilon^2-x^2} \cdot \overline{B L^{-1}_\varepsilon u_{\theta,\varepsilon}(x)}\,
\frac{u_{\theta,\varepsilon}(x)}{\sqrt{\varepsilon^2-x^2}}dx\Big|\\
\le  \|B\|_{L(X_\varepsilon,X_\varepsilon)} \cdot \|L^{-1}_\varepsilon u_{\theta,\varepsilon}\|_{X_\varepsilon}
\cdot
\sqrt{\int_{-\varepsilon}^\varepsilon 
\frac{\big|u_{\theta,\varepsilon}(x)\big|^2}{\sqrt{\varepsilon^2-x^2}}dx}\\
\le a  \|B\|_{L(X_\varepsilon,X_\varepsilon)} \|L^{-1}_\varepsilon u_{\theta,\varepsilon}\|_{X_\varepsilon},
\quad a>0,
\end{eqnarray*}
and the previous estimates show that $\|L^{-1}_\varepsilon u_{\theta,\varepsilon}\|_{X_\varepsilon} = O(1/\log\varepsilon)$.
Proposition \ref{prop2} is proved.

\medskip

Let us conclude the proof of theorem \ref{thm1}.
Assume  that $K_{\theta,\varepsilon}(z)f_{\theta,\varepsilon}(z)=0$ with $f_{\theta,\varepsilon}\ne 0$
and $z=E(\theta,\varepsilon)$ sufficiently close to $E$.
Using the invertibility of $L_\varepsilon$ and the decomposition (\ref{eq-decomp})
one obtains
\[
\Big(1 - 2\pi L^{-1}_\varepsilon R_{\theta,\varepsilon}\Big) f_{\theta,\varepsilon}=
\frac{2\pi}{E-z} \,\langle u_{\theta,\varepsilon}, f_{\theta,\varepsilon}\rangle L^{-1}_\varepsilon u_{\theta,\varepsilon}.
\]
The operator $1 - 2\pi L^{-1}_\varepsilon R_{\theta,\varepsilon}$ is invertible for $\varepsilon$ small enough, therefore,
\[
 f_{\theta,\varepsilon}=\frac{2\pi}{E-z}\, \big(1 - 2\pi L^{-1}_\varepsilon R_{\theta,\varepsilon}\big)^{-1} \langle u_{\theta,\varepsilon}, f_{\theta,\varepsilon}\rangle L^{-1}_\varepsilon u_{\theta,\varepsilon},
\]
which implies $\langle u_{\theta,\varepsilon}, f_{\theta,\varepsilon}\rangle\ne 0$ (otherwise one would get $f_{\theta,\varepsilon}\equiv 0$).
Taking the scalar product $\langle\cdot,\cdot\rangle$ of the both parts with $u_{\theta,\varepsilon}$
one arrives at
\[
z=E(\theta,\varepsilon)=E-2\pi \Big\langle
\big(1 - 2\pi L^{-1}_\varepsilon R_{\theta,\varepsilon}\big)^{-1}  L^{-1}_\varepsilon u_{\theta,\varepsilon},
u_{\theta,\varepsilon}\Big\rangle.
\]
Using (\ref{eq-oeps}), let us fix $B>0$ such that
\begin{equation}
   \label{eq-eps2}
\|L^{-1}_\varepsilon R_{\theta,\varepsilon}(z)\|_{L(X_\varepsilon,X_\varepsilon)}\le \frac{B}{|\log \varepsilon|}
\end{equation}
for small $\varepsilon$. As the map $f\mapsto \langle f, u_{\theta,\varepsilon}\rangle$ is
continuous in the $X_\varepsilon$-norm, we have,
for sufficiently small $\varepsilon$ such that $\frac{2\pi B}{|\log \varepsilon|}<1$,
\begin{eqnarray*}
\Big\langle
\big(1 - 2\pi L^{-1}_\varepsilon R_{\theta,\varepsilon}\big)^{-1}  L^{-1}_\varepsilon u_{\theta,\varepsilon},
u_{\theta,\varepsilon}\Big\rangle\qquad\qquad\qquad\qquad\qquad\qquad\\
{}=\big\langle L^{-1}_\varepsilon u_{\theta,\varepsilon},
u_{\theta,\varepsilon}\big\rangle
+\sum_{k\ge 1}
\Big\langle
\big(2\pi L^{-1}_\varepsilon R_{\theta,\varepsilon}\big)^{k}  L^{-1}_\varepsilon u_{\theta,\varepsilon},
u_{\theta,\varepsilon}\Big\rangle=:D_1+D_2.
\end{eqnarray*}
In view of the estimate (\ref{eq-est1}) for $D_1$, it is sufficient show that $D_2=O\big(1/|\log\varepsilon|^2\big)$.
By (\ref{eq-oeps}) and (\ref{eq-est2}),
\[
\Big|\Big\langle \big(2\pi L^{-1}_\varepsilon R_{\theta,\varepsilon}\big)^{k}  L^{-1}_\varepsilon u_{\theta,\varepsilon},
u_{\theta,\varepsilon}\Big\rangle\Big|
\le \frac{A}{|\log\varepsilon|}\Big(
\frac{2\pi B}{|\log\varepsilon|}\Big)^k,
\]
hence
\[
|D_2|\le \frac{A}{|\log\varepsilon|}
\sum_{k\ge 1 } \Big(
\frac{B}{|\log\varepsilon|}\Big)^k=
\frac{A}{|\log\varepsilon|} \cdot \frac{\frac{2\pi B}{|\log\varepsilon|}}{1-\frac{2\pi B}{|\log\varepsilon|}}
=O(1/|\log\varepsilon|^2),
\]
which finishes the proof of theorem \ref{thm1}.

\end{document}